\documentstyle[twocolumn,aps,epsfig]{revtex}
\begin{document}
\wideabs{
\draft
\title{\bf{\Large{Stationary localized states due to nonlinear impurities
described by the modified discrete nonlinear Schr\"odinger equation }}}
\author{Bikash C. Gupta and Sang Bub Lee}
\address{Department of Physics, Kyungpook National University, Taegu 
702-701, Korea}
\maketitle
\begin{abstract}
The modified discrete nonlinear Schr\"odinger equation is used to 
study the formation of stationary localized states in a one-dimensional 
lattice with a single impurity and an asymmetric dimer impurity. 
A periodically modulated and a perfectly nonlinear chain is also 
considered.
Phase diagrams of localized states for all systems are presented. 
From the mean square displacement calculation, it is found that 
all states are not localized even though the system comprises
random nonlinear site energies. Stability of the states is discussed. 
\end{abstract}

\narrowtext
}
\section{Introduction}
Localized states appear due to the presence of impurity or disorder (which 
breaks the translational symmetry) in the system \cite{econ}. While there 
have been studies on the formation of localized states due to linear 
impurities in various systems \cite{econ}, only a few look into the 
formation of localized states due to nonlinear impurities. The discrete 
nonlinear Schr\"odinger equation [2-16] employed to study the formation 
of stationary localized (SL) states is given by  
\begin{equation}
i\frac{dC_n}{dt}=\epsilon_n C_n + V(C_{n+1} + C_{n-1}) 
- \chi_n |C_n|^{\sigma} C_n,
\end{equation}
where $\sigma$ is a constant,  $C_n$ is the probability amplitude 
of the particle (exciton) to be at the site $n$, $\epsilon_n$ and $\chi_n$ 
are, respectively, the static site-energy and the nonlinear strength at 
site $n$, and $V$ is the 
nearest neighbor hopping element. The nonlinear term $|C_n|^\sigma C_n$
arises due to the interaction of the exciton with the lattice 
vibration \cite{ken1,ken2}. If the oscillators are purely harmonic in 
nature, $\sigma = 2$. The Eq. (1), has been used to study 
the formation of SL states in a one-dimensional chain as well as in the 
Cayley tree with single and dimer nonlinear impurities
\cite{mol1,mol2,mol3,hui1,hui2,bik2,bik3,kund}. For the case of a
perfectly nonlinear chain where $\chi_n=\chi$ for all $n$ it has been 
shown that 
SL states are possible even though the translational symmetry of
the system is preserved \cite{acev,wein,kund,bik1,ghos}. 
These results were remarkably different from those of the corresponding 
systems with linear impurities. 
The Eq. (1) has been derived with the assumption that the lattice 
oscillators in the system are local and their motions are much faster than
the exciton motion. A natural question to be raised is what happens to the 
formation of SL states when the oscillators in the lattice are harmonic
in nature and coupled with their nearest neighbors. In such system, in the
presence of the interaction of the exciton with the lattice vibration, the
time evolution equation governing the dynamics of the exciton has been 
derived by Kopidakis {\it et al.} \cite{kopi} as 
\begin{eqnarray}
i\frac{dC_n}{dt}=&\epsilon_n& C_n + V(C_{n+1} + C_{n-1}) \nonumber \\
&-& \chi_n (|C_{n+1}|^2
+|C_{n-1}|^2 + 2|C_n|^2)C_n,
\end{eqnarray}
where $C_n$, $\epsilon_n$, $V$ and $\chi_n$ carries the same meanings as 
in Eq.~(1). The Eq.~(2) is called the {\it `modified'} discrete nonlinear 
Schr\"odinger equation. The Hamiltonian which generates Eq.~(2) is
given by 
\begin{eqnarray}
H&=&\sum_{n=-\infty}^{\infty} \left(C_n^\star C_{n+1} + C_nC_{n+1}^\star 
\right) \nonumber \\
&+&\sum_{n=-\infty}^\infty\frac{\chi_n}{2} \left(|C_{n+1}|^2 + |C_{n-1}|^2 
+2 |C_n|^2 \right)|C_n|^2.
\end{eqnarray}
We note that the Eq. (2) has more nonlinear 
terms compared to the Eq.~(1) and, in addition, the Eq.~(2) is more relevant 
in condensed matter physics. To the best of our knowledge, this equation 
has not been used to study the formation of SL states even though an 
equation similar to Eq.~(2) has been used to study the formation SL 
states in one-dimensional system due to a single impurity and a symmetric 
dimer impurity only \cite{gupt}. Therefore, it is essential to consider the 
right equation (i.e., Eq. (2)) for the elaborate study of the 
formation of SL states in one-dimensional systems. Thus we study it here.

This paper is organized as follows. In Sec.~II, we discuss the
effect of a single impurity (i.e., $\chi_n=\chi \delta_{n,0}$)
on the formation of SL states. In Sec.~III we consider the case of 
a dimer impurity, $\chi_n = (\chi_1 \delta_{n,0} + 
\chi_2 \delta_{n,1})$. In Sec.~IV the periodically modulated nonlinear 
chain as well as a perfectly nonlinear chain will be considered. 
In Sec.~V, we consider the presence of random nonlinearities through out 
the lattice. The stability of the states are discussed in Sec.~VI.
Finally we summarize our findings in Sec.~VII.

\section{single nonlinear impurity}
Consider the system of a one-dimensional chain with a nonlinear impurity 
at the cener site. The time evolution of an exciton in the system is 
governed by Eq.~(2) with $\chi_n = \chi \delta_{n,0}$. The Hamiltonian
which produces the equation of motion for the exciton in the system 
is given by
\begin{eqnarray}
H&=&\sum_n (C_n^\star C_{n+1} + C_n C_{n+1}^\star) \nonumber \\
&-& \frac{\chi}{2} \left(|C_1|^2
+ |C_{-1}|^2 + 2 |C_0|^2 \right) |C_0|^2.
\end{eqnarray} 
Since $\sum_n |C_n|^2$ is a constant of motion, we suitably renormalize 
so that $\sum_n |C_n|^2$ = $1$. Therefore, $|C_n|^2$ can be considered 
as the probability for the exciton to be at site $n$.  Since 
we are interested in the stationary localized states, we consider
the ansatz
\begin{equation}
C_n=\phi_n \exp(-iEt);~~~~~\phi_n=\phi_0 \eta^{|n|}
\end{equation}
where $\eta$ lies between 0 and 1 and can be asymptotically defined as 
$\eta = \frac{|E|-\sqrt{E^2-4}}{2}$, $E$ being the energy of the localized 
state which appears outside the host band. Since in a one-dimensional 
system, states
appearing outside the band at the expanse of the states in the continuum 
are exponentially localized, the ansatz in Eq. (5) is justified and may 
also be readily derived from the Green function analysis \cite{bik2,bik3}. 
Substituting the ansatz to the normalization condition we get
\begin{equation}
|\phi_0|^2 = \frac{1-\eta^2}{1+\eta^2}.
\end{equation}
Direct substitution of $\phi_0$, $\phi_n$ and hence $C_n$ in terms of
$\eta$ in Eq. (4) yields an effective Hamiltonian,
\begin{equation}
H_{{\rm eff}} = \frac{4\eta}{1+\eta^2} - \frac{\chi (1-\eta^2)^2}{(1+\eta^2)}.
\end{equation}
The fixed point solutions of the reduced dynamical system described by 
$H_{{\rm eff}}$ will give the values of $\eta$ which correspond to 
the localized state solutions \cite{wein}. It should be noted that 
since $\chi$ is constant, the effective Hamiltonian is a function of 
only one dynamical variable $\eta$. 
Thus fixed point solutions are readily obtained from the condition
$\partial{H_{{\rm eff}}} /\partial{\eta}$ = 0, {\em i.e.},
\begin{equation}
\frac{2}{|\chi|}=\eta (3+\eta^2) =f(\eta).
\end{equation}
Thus, for a fixed value of $\chi$, the number of solutions of Eq.~(8) lying 
between 0 and 1 will correspond to the number possible SL states. We note
that $f(\eta)$ is a monotonically increasing function of $\eta$ and will be 
maximum at $\eta = 1$. Thus, we see that there is no solution for $|\chi| 
< 0.5$ and only one solution for $|\chi| \ge 0.5$. Therefore, we conclude 
that there is a critical value of $|\chi| = 0.5$ separating the no-state 
region from the one-state region. On the other hand, for the same system 
described by Eq.~(1), the critical value of $|\chi|$ separating the 
no-state region from the  one-state region is 2. Thus the nearest neighbor 
coupling of the lattice oscillators are responsible to reduce the critical 
value of nonlinear strength by a factor of four. Therefore, we expect that
next nearest neighbor coupling of the lattice oscillators will further 
reduce the critical value of the nonlinear strength and thus the 
critical value may approach to zero when the oscillators are globally 
coupled.

\section{Dimer Nonlinear Impurity}
We consider a one-dimensional lattice with two nonlinear impurities placed
at site 0 and 1 denoted by $\chi_n$ = ($\chi_1 \delta_{n,0}$ 
+ $\chi_2 \delta_{n,1}$). If $\chi_1 = \chi_2$, we call it symmetric-dimer 
impurity and otherwise asymmetric-dimer impurity. We will here consider the 
system with asymmetric-dimer impurity, since the results for the  
symmetric-dimer impurity will be obtained as a special case. The 
one-dimensional chain with 
the asymmetric dimer impurity can be described by Eq.~(3) with $\chi_n$ = 
($\chi_1 \delta_{n,0}$ + $\chi_2 \delta_{n,1}$). For stationarity condition 
we assume that $C_n = \phi_n \exp(-iEt)$. Furthermore, for localized states 
we assume the following forms for $\phi_n$.
\begin{eqnarray}
\phi_n = [{\rm sgn}(E) \eta]^{n-1} \phi_1 ; ~~~~~~~~~~~~~~n \ge 1, \nonumber \\
\phi_{-|n|} = [{\rm sgn}(E) \eta]^{|n|} \phi_0 ; ~~~~~~~~~~~~n \le 0,
\end{eqnarray}
with $\eta$ as defined earlier. The ansatz given by Eq.~(9) is exact and
can be derived from the Green function analysis. It is also 
justified from the fact that the states which appear outside the host 
band are exponentially localized \cite{bik2}. Three different cases 
are possible: (i) $\phi_1 = \phi_0$ (symmetric), 
(ii) $\phi_1 = -\phi_0$ (antisymmetric) and (iii) $\phi_1 \ne \phi_0$ 
(asymmetric). To consider all three cases, we introduce a variable
$\beta = \frac{\phi_0}{\phi_1}$. Without any loss of generality,
$\phi_0$ and $\phi_1$ can be assumed to be real. Therefore, the value of 
$\beta$ is confined between 1 and $-1$ if $|\phi_0| \le |\phi_1|$. Else we 
inverse the definition of $\beta$. Thus,  $\beta = 1$ corresponds to the
symmetric states, $\beta = -1$ corresponds to the antisymmetric states, 
and $-1 \le \beta \le 1$ corresponds to the asymmetric states. 
Substituting the ansatz in Eq.~(9) and the definition of $\beta$ 
in the normalization condition, $\sum_n|C_n|^2 = 1$, we obtain
\begin{equation}
|\phi_0|^2 = \frac{1-\eta^2}{1+\beta^2}.
\end{equation}  
Using the ansatz and the normalization condition, we obtain the effective
Hamiltonian as a function of $\beta$ and $\eta$, given as 
\begin{eqnarray}
H_{{\rm eff}}&=&2 \beta \frac{1-\eta^2}{1+\beta^2} + 2 {\rm sgn}(E) \eta 
\nonumber \\
&-& \frac{\chi_1}{2} \left(\frac{\left(1-\eta^2\right)^2}{1+\beta^2}  
+ \frac{\left(1+\eta^2\right) \left(1-\eta^2\right)^2}
{\left(1+\beta^2\right)^2} \right) \nonumber \\
&-& \frac{\chi_2}{2} \left(
\frac{\left(\beta^2+\beta^4\eta^2+2\beta^4\right)\left(1-\eta^2\right)^2}
{\left(1+\beta^2\right)^2}  \right).
\end{eqnarray}
If $\beta = \pm 1$, we obtain 
\begin{eqnarray}
H_{{\rm eff}}^{\pm} =& & \pm (1 - \eta^2) + 2 {\rm sgn}(E) \eta 
\nonumber \\
&-& \frac{\chi_1}{2}
\left(\frac{\left(1-\eta^2\right)^2}{2} + \frac{\left(1+\eta^2\right) 
\left(1-\eta \right)^2}{4} \right) \nonumber \\
&-&\frac{\chi_2}{2}
\left(\frac{\left(3+\eta^2\right) \left(1-\eta^2 \right)^2}{4}\right)  
\end{eqnarray}
where `$+$' corresponds to the symmetric states and `$-$' corresponds to
the antisymmetric states. The number of  fixed point solutions of the 
reduced dynamical system described by $H_{{\rm eff}}$ gives the possible 
number of SL states. The fixed point solutions satisfy the equation,   
\begin{eqnarray}
&4&\left({\rm sgn}(E)\mp\eta\right) \nonumber \\
&+&\chi_1\left(2\eta\left(1-\eta^2\right)-
\frac{\eta\left(1-\eta^2\right)^2}{2} 
+ \eta\left(1-\eta^4\right)\right) 
\nonumber \\  
&+&\chi_2 \left(\eta\left(1-\eta^2\right) \left(3+\eta^2\right)-
\frac{\eta\left(1-\eta^2\right)^2}{2} \right) = 0.
\end{eqnarray}
~~
\vspace{.5cm}
     \begin{center}
      \epsfig{file=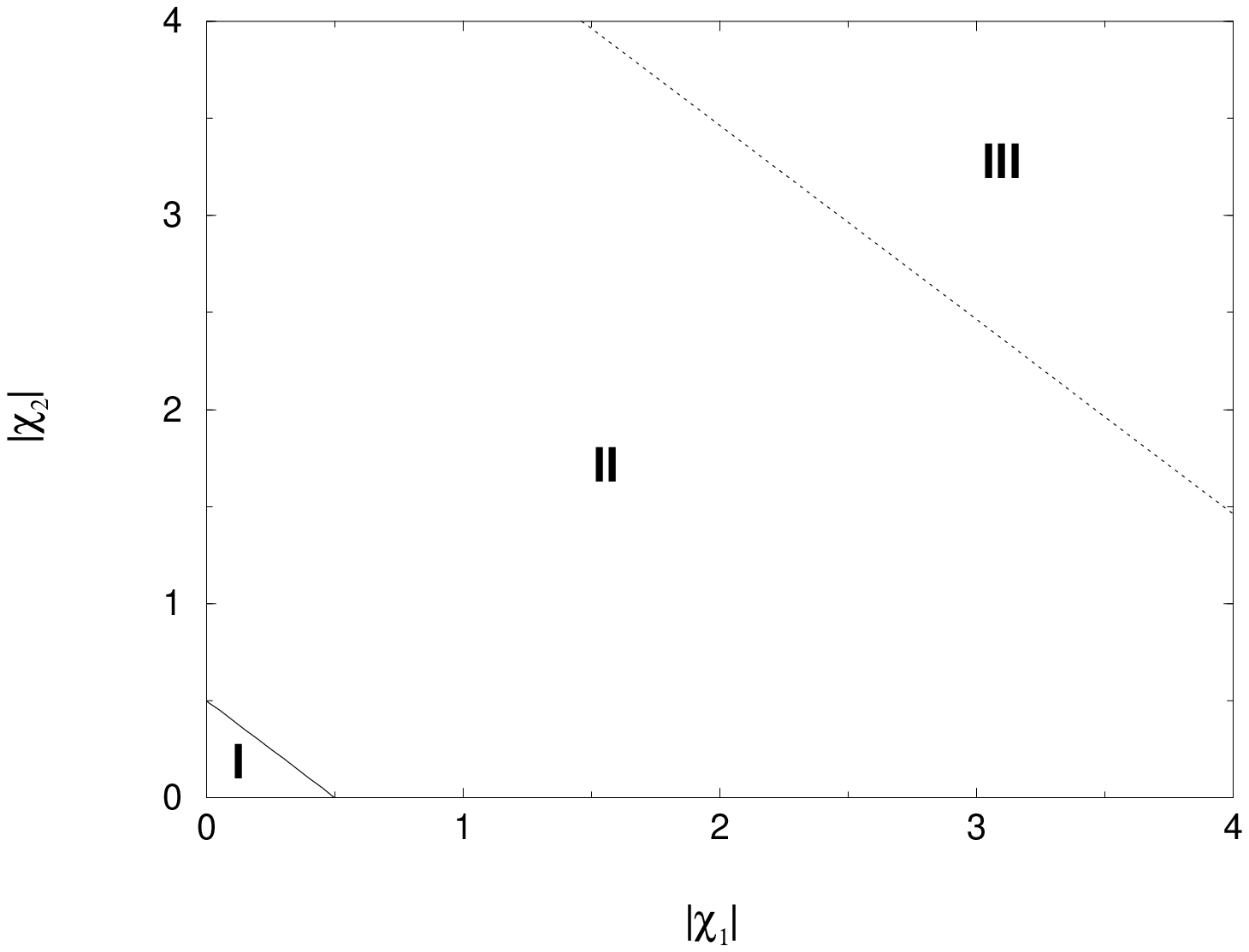,width=7.4cm,height=5.2cm}
      \vspace{0.5cm}
      \begin{figure}[p]
      \caption{Phase diagram of symmetric and antisymmetric SL states for
         the one-dimensional chain with a dimer impurity. Region I
         has no SL state, region II has one symmetric SL state and region III
         has one symmetric SL state and two antisymmetric SL states. 
      \label{fig1}}
      \vspace{0.5cm}
      \end{figure}
      \end{center}
After solving this equation for each point in the $(|\chi_1|,|\chi_2|)$ 
plane, we obtain the number of possible symmetric as well as antisymmetric 
SL states in the $(|\chi_1|,|\chi_2|)$ plane. 
The phase diagram is shown in Fig. 1. The solid line is the critical line
separating the no-symmetric-state region from the one symmetric-state-region 
and, the dotted line is the critical line separating no-antisymmetric-state 
region from the two-antisymmetric-states region. Thus, 
there is no state in region I, one symmetric state in region II, and one 
symmetric and two antisymmetric states in region III.
Furthermore, there is one symmetric state on the solid and, one symmetric 
state and one antisymmetric state on the the dotted line. 

Now, let us find the asymmetric states where $\beta \ne 1$. The
effective Hamiltonian is a function of two dynamical variables,
$\beta$ and $\eta$. Therefore the fixed point solutions will obey the
equations given by 
\begin{equation}
\frac{\partial{H_{{\rm eff}}}}{\partial\eta} = 0 ~~{\rm and} ~~\frac{\partial
H_{{\rm eff}}}{\partial{\beta}} = 0.
\end{equation} 
After a little algebra we obtain the desired equations,
\begin{eqnarray}
&2&\left(1-\beta^4 \right)+\chi_1 \beta \left(1-\eta^2 \right) 
\left(3+\beta^2+2\eta^2\right)  
\nonumber \\
&-&\chi_2 \beta\left(1-\eta^2\right)
\left(1+2\beta^2\eta^2+3\beta^2\right) = 0, \\
&{\rm and}& \nonumber \\
&4&\beta\eta\left(1+\beta^2\right)-2\left(1+\beta^2 \right)^2  -\chi_1\eta
\left(1-\eta^2\right) \left(3+2\beta^2+3\eta^2 \right) 
\nonumber \\
&-&\chi_2 \eta \left(1-\eta^2\right)
\left(2\beta^2 + 3\beta^4\eta^2 + 3\beta^4\right) = 0. 
\end{eqnarray}
~~
\vspace{0.5cm}
     \begin{center}
      \epsfig{file=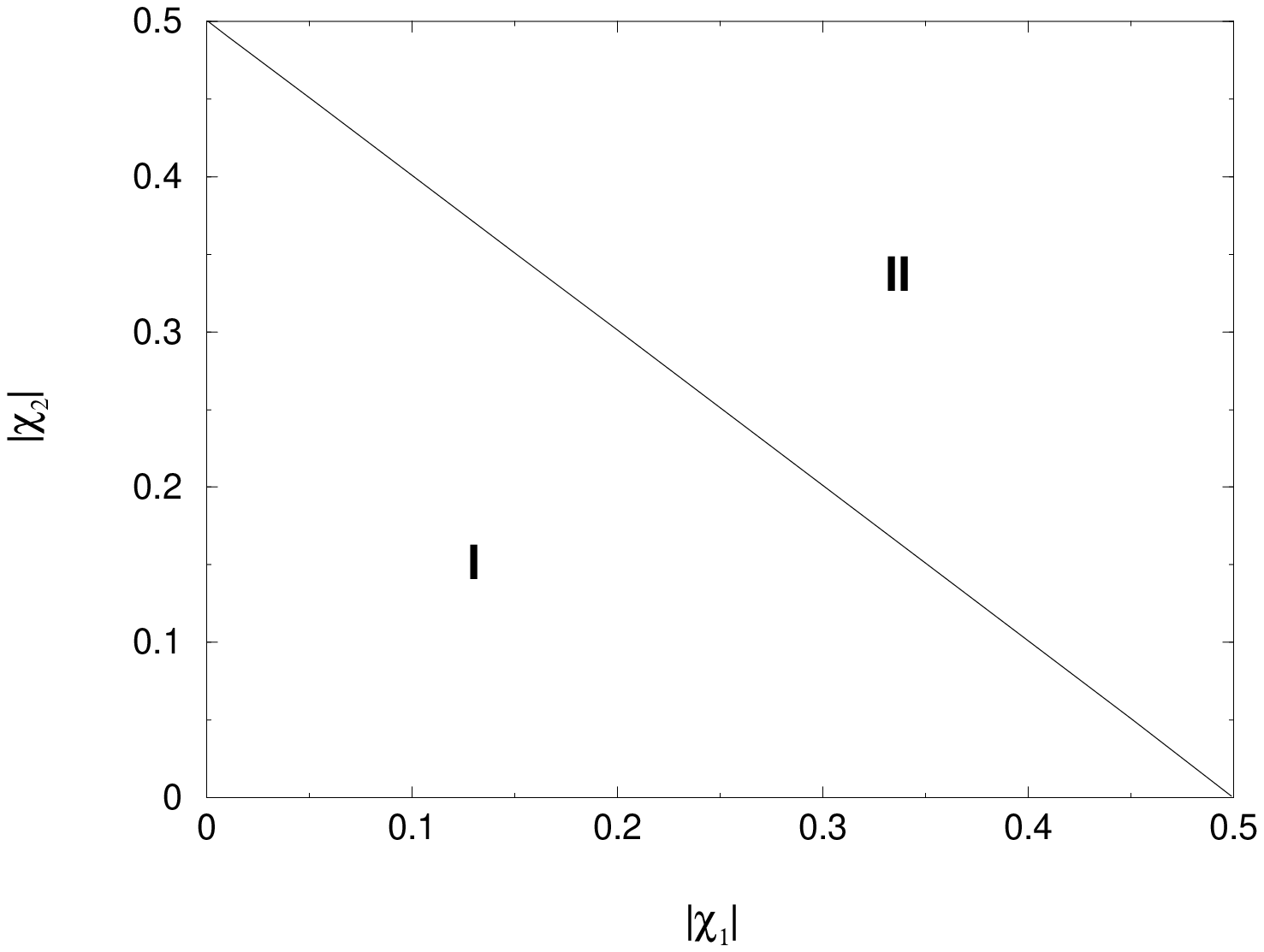,width=7.4cm,height=5.2cm}
      \vspace{0.3cm}
      \begin{figure}[p]
      \caption{Phase diagram of asymmetric SL states for
         the one-dimensional chain with a asymmetric-dimer impurity.
         Solid line is the critical line separating no-asymmetric-state
         region (marked as I) from the one-asymmetric-state region
         (marked as II). 
      \label{fig2}}
      \vspace{0.5cm}
      \end{figure}
      \end{center}
Plotted in Fig.~2 is the phase diagram obtained from the solutions of Eqs.~
(15) and (16). There is only one
critical line (for $|\chi_1| \ne |\chi_2|$) separating the no-asymmetric-state 
region (marked as I) from the one-asymmetric-state region (marked as II). 
The critical line in Fig.~2, in fact, turns out to be very closed to 
the solid critical line in Fig.~1. However, 
$|\chi_1|$ = $|\chi_2|$ is a special case as far as asymmetric SL state 
is concerned. For $|\chi_1|$ = $|\chi_2|$, the critical value of $|\chi|$
separating the no-state from the one-state region is 2.015.   
The full phase diagram of SL states for the dimer impurity may be obtained
by superposing Fig.~1 and Fig.~2 (not shown).  
It is clear from the Fig.~1 and Fig.~2 that there are three distinct 
regions having one, two, and four states in the $(|\chi_1|, |\chi_2|)$ plane. 
Furthermore we see that there is a continuity as we go from
one region to another. For the symmetric dimer ($|\chi_1|=|\chi_2|=|\chi|$), 
we obtain three critical points, i.e., $|\chi_{cr}^{(1)}| = 0.25$, 
$|\chi_{cr}^{(2)}|$ = 2.015 and $|\chi_{cr}^{(3)}|= 2.732$. There is no 
SL state for $|\chi| < |\chi_{cr}^{(1)}|$, one state for $|\chi_{cr}^{(1)}| 
\le |\chi| < |\chi_{cr}^{(2)}|$, two  states for $|\chi_{cr}^{(2)}| \le 
|\chi| < |\chi_{cr}^{(3)}|$, three  states for $|\chi| = |\chi_{cr}^{(3)}|$
and four  states for $|\chi| > |\chi_{cr}^{(3)}|$. The corresponding critical 
values obtained by Eq.~(1) are 1, 8/3 and 8 respectively. 
Thus, in case of symmetric dimer also it is observed that the nearest neighbor 
coupling of the lattice oscillators is responsible to reduce the critical 
values, while the maximum number of SL states remain unchanged. 
However, it should be noted that the 
asymmetric nature of the dimer does play a crucial role to reduce the 
critical value of nonlinear strength abruptly.  

\section{Fully Nonlinear Chain}
We now consider a periodically modulated nonlinear chain, i.e., the 
alternate sites are of the same nonlinear strength but any two consecutive
sites are of the different strength. The distribution of the nonlinear 
strength,
$\chi_n$ is given by
\begin{equation} 
\chi_{2n+1}=\chi_1;~~ \chi_{2n}=\chi_2;~~ n= -\infty,...-1,0,1,
...\infty. 
\end{equation} 
When $\chi_1=\chi_2$, the system reduces to a perfectly nonlinear chain.
In both cases the translational symmetry of the system remains preserved.
The Hamiltonian for this system is given by Eq.~(3) with $\chi_n$ as defined 
in Eq.~(17). Using the stationarity condition, $C_n=\phi_n \exp(-iEt)$, we can 
obtain the Hamiltonian in terms of $\phi_n$. In this case it is not possible 
to find the exact ansatz for the localized states, but there are a few 
rational choices. For example, the on-site peaked as well as the inter-site 
peaked and dipped solutions are possible. We will consider these cases
subsequently.

\subsection{On-site peaked solution}
Since the lattice is periodically modulated with a period of two, the 
localized solution may have peak either at a site having nonlinear strength 
$\chi_1$ or at a
site having strength $\chi_2$. We will carry out the calculation assuming 
that the solution is peaked at the center site (site 0).
The result for the solution peaked at first site (site 1) may easily be 
obtained by interchanging the two nonlinear
strengths in the resulting expression for the case of the center site peaked 
solution.
Therefore, using the ansatz $\phi_n = \phi_0 \eta^{|n|}$  
 and the normalization condition we get the effective Hamiltonian, 
\begin{eqnarray}
H_{{\rm eff}}&=&\frac{4\eta}{1+\eta^2} 
- \frac{\chi_2}{2} \left(\frac{2\eta^2\left( 
1-\eta^2\right)}{\left(1+\eta^2\right)^3} + \frac{2\left(1-\eta^2\right)
\left(1+\eta^8\right)}{\left(1+\eta^2\right)^3 
\left(1+\eta^4\right)}\right) \nonumber \\
&-& \frac{\chi_1}{2} \left(\frac{2\eta^2\left(1-\eta^2\right)}{\left(
1+\eta^2\right)^3}+\frac{4\eta^4\left(1-\eta^2\right)}
{\left(1+\eta^2 \right)^3 \left(1+\eta^4\right)} \right).
\end{eqnarray}
After analyzing the fixed point equation, $\partial{H_{{\rm eff}}}/
\partial{\eta}$ = 0, we find that there is only one SL state. 
Interchanging $\chi_1$ with $\chi_2$, we obtain another SL state.
Therefore for fixed value of $\chi_1$ and $\chi_2$, we will have two
SL states, one peaked at a site of nonlinear strength $\chi_1$ and 
another peaked at a site of nonlinear strength $\chi_2$. 
Thus we conclude that if the nonlinear chain is modulated with periodicity 
$n$, there will be $n$ number of on-site peaked localized solutions. 
We note that the extra SL state due to extra periodicity in the periodically 
modulated chain was overlooked when the same system was studied by the use 
of Eq.~(1) \cite{ghos}. These two states get closer and become the 
same as $\chi_1 \rightarrow \chi_2$. After substituting 
$\chi_1$=$\chi_2$=$\chi$ in the 
fixed point equation we obtain
\begin{equation}
\frac{1}{|\chi|} = \frac{\eta}{1-\eta^2}.
\end{equation}
Equation (19) clearly indicates that there is one on-site site peaked 
solution (SL state) for a perfectly nonlinear chain, irrespective of the 
strength of the nonlinearity.

\subsection{Inter-site peaked and dipped solution}
For the inter-site peaked and dipped solutions we use the dimer ansatz
in Eq. (9). After some algebra we obtain the effective Hamiltonian for the 
reduced dynamical system as
\begin{eqnarray}
H_{{\rm eff}} = 2 {\rm {\rm sgn}}(E)\eta +\frac{2\beta\left(1-\eta^2\right)}
{\left(1+\beta^2 \right)} - 
{\widetilde\chi}\left(\frac{\beta^2(1-\eta)^2}
{(1+\beta^2)^2}\right)\nonumber \\
+{\widetilde\chi}\left(\frac{\eta^6(1-\eta^2)^2(1+\beta^4)}
{(1+\beta^2)^2(1-\eta^8)} + \frac{\eta^2(1-\eta^2)^2(1+\beta^4)}
{(1+\beta^2)^2(1-\eta^8)} \right) \nonumber \\
-\frac{\chi_1}{2}\left(\frac{2(1-\eta^2)^2(1+\beta^4\eta^4)}
{(1+\beta^2)^2 (1-\eta^8)} \right) 
\nonumber \\
- \frac{\chi_2}{2}\left( 
\frac{2(1-\eta^2)^2(\beta^4+\eta^4)}{(1+\beta^2)^2(1-\eta^8)}\right) 
\end{eqnarray}
where ${\widetilde\chi}$=$\chi_1$
+$\chi_2$. We first consider the case, $\beta = \pm 1$. Substituting $\beta 
= \pm 1$ to the Eq.~(20) we obtain 
\begin{equation}
\frac{4}{|{\widetilde \chi^{\pm}}|}= \frac{\eta (2+\eta^2-\eta^4)}
{(1 \mp \eta)(1+\eta^2)}
\end{equation}
from the fixed point equation.
Here the `$+$' corresponds to the symmetric SL states and the `$-$'  to the 
antisymmetric SL states.
From Eq.~(21) it is obvious that there will always be one symmetric solution
for any point in the $(|\chi_1|,|\chi_2|)$ plane. But for 
antisymmetric solutions there are two critical lines. One being 
$|\chi_2|$=6.427 $-$ $|\chi_1|$ ( dotted line in Fig. 3) and another 
$|\chi_2|$ = 8 $-$ $|\chi_1|$ (solid line in Fig. 3). 
Three distinct regions (in Fig.~3) separated by the critical lines 
are marked by I, 
II and III. Region I has no SL state, region II has two antisymmetric 
states and region III has only one antisymmetric  state. Furthermore 
we note that, for $\chi_1 = \chi_2=\chi$, there is no antisymmetric 
state for $|\chi| < 3.2135$,  
two states for $3.2135 < |\chi| \le 4$, and one solution for $|\chi| > 4$. 
These critical values are again lower than those obtained from Eq. (1). 
\vspace{0.5cm}
     \begin{center}
      \epsfig{file=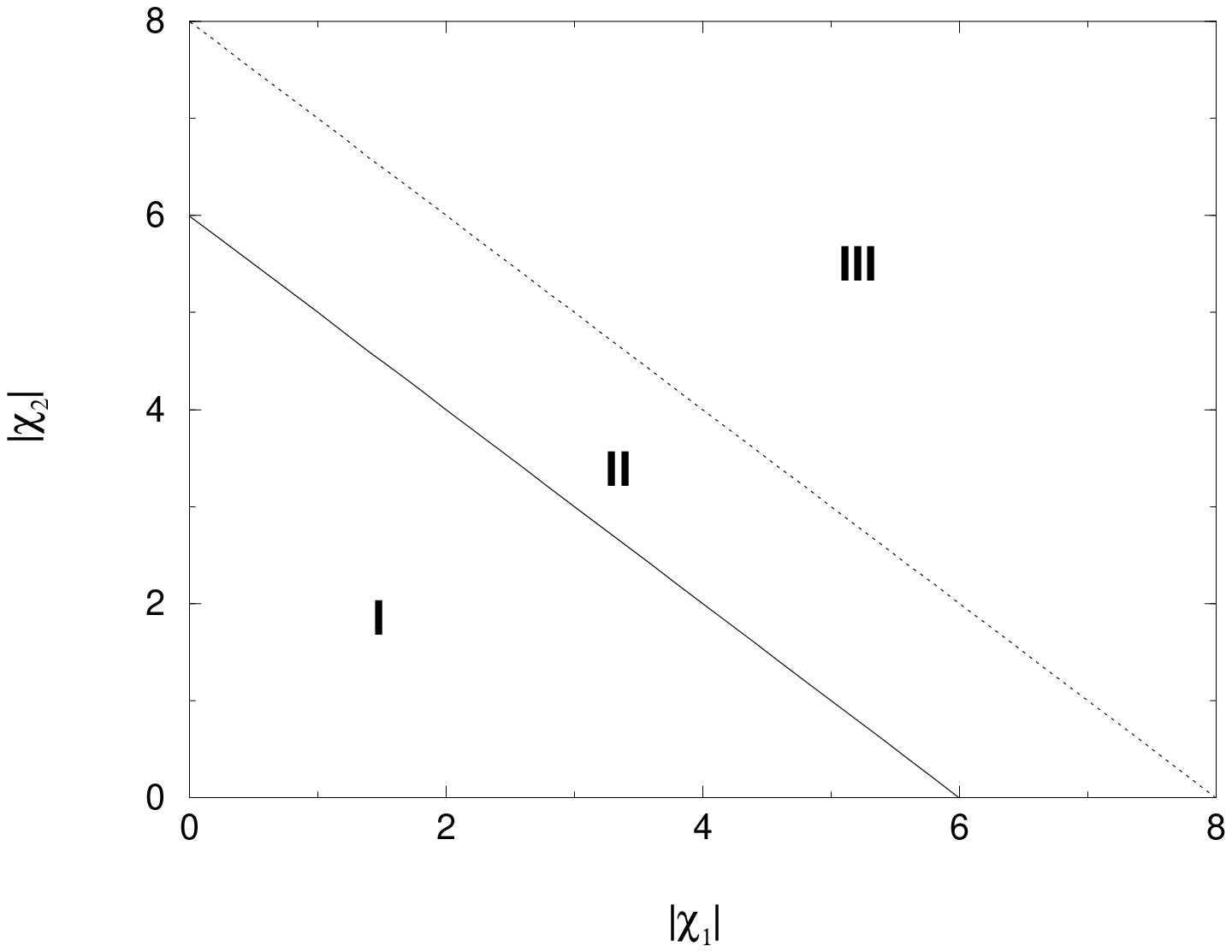,width=7.4cm,height=5.2cm}
      \vspace{0.4cm}
      \begin{figure}[p]
      \caption{Phase diagram of SL states for a periodically modulated 
               nonlinear chain. The regions I, II and III has no, two and 
	       one antisymmetric SL state respectively. 
      \label{fig3}}
      \vspace{0.5cm}
      \end{figure}
      \end{center}
Now, we look for the asymmetric solution corresponding to $\beta \ne1$.
The fixed-point equations are lengthy and, therefore, are not presented 
here. The fixed-point equations are solved numerically and  the possible    
number of asymmetric SL states in the $(|\chi_1|,|\chi_2|)$ plane are
found. It turns out that there is            
always one asymmetric SL state in the $(|\chi_1|,|\chi_2|)$ plane except
along the line $|\chi_1| = |\chi_2|$ on which there exists a critical point,
$|\chi_1| = |\chi_2| = 1.75$, below which there is no asymmetric SL state
and above which one asymmetric SL state. Thus, it is found that 
the modulation of nonlinearity  reduces the critical value to obtain a 
localized state.

We emphasize that the SL states obtained in the fully nonlinear systems
are based on the rational ansatz. However, many other SL states
may possibly exist in the system.
Thus, we are not in a situation to conclude about the maximum number
of SL states for the fully nonlinear chain.

\section{Random nonlinearity}
Here we consider a fully nonlinear chain on which all sites  have
random nonlinear strengths. It is known that, due to static random site
energies, all the states in the system are localized and the system
behaves as an
insulator. The question is whether or not all the states are localized
due to random nonlinearities in the system. Since the appropriate ansatz
is not known, we calculate the mean square displacement of a particle
in the system consisting of random nonlinear site energies \cite{molina}.
The evolution of exciton is defined by Eq. (2) with $\chi_n$ being chosen
randomly with equal probability for all $n$. The mean square displacement
is defined as
\begin{equation}
m(t)=\sum_{i=-\infty}^{\infty} i^2 |C_i(t)|^2.
\end{equation}
    \begin{center}
      \epsfig{file=fig4.eps,width=7.4cm,height=5.2cm}
      \vspace{0.4cm}
      \begin{figure}[p]
      \caption{The $<m(t)>/t^2$ is plotted as a function of time $t$. 
      \label{fig4}}
      \vspace{0.5cm}
      \end{figure}
      \end{center}
Since the set of coupled equations [Eq.~(2)] can not be solved 
analytically, we
solve them numerically using the fourth order Runge-Kutta method.
Self-expanding
lattice is used to avoid the finite size effect. The particle is
initially placed at site 0 i.e., the center site. The time
interval for the calculation is taken to be 0.001. The total probability
of the particle in the system should be conserved and it has been confirmed
at every time step of our simulation. The mean square displacement is
averaged over hundred random configurations and denoted as $<m(t)>$.
Shown in Fig. 4 is $<m(t)>/t^2$ as a function of $t$.  We see that
$<m(t)>/t^2$ initially decreases and becomes constant, much smaller
than 2. Thus, we conclude that there is a partial ballistic flow of the
particle and hence all the states are not localized unlike the case
of a linear system with  random site energies. 

\section{Stability}
The stability of the SL states can be understood from a flow diagram.
For this purpose, we consider antisymmetric solutions due to the presence
of a symmetric dimer in one dimensional chain. The reduced dynamical
Hamiltonian for this case is given by
\begin{equation}
H_{{\rm eff}}=-(1-\eta^2)+2\eta +\frac{|\chi|}{4}((3+\eta^2)(1-\eta^2)^2).
\end{equation}
The flow diagram of the dynamical system described by the $H_{{\rm eff}}$
is constructed in the following manner. We treat $\partial{H_{{\rm eff}}}
/\partial{\eta}$ = $f(\eta)$ as the velocity and $\eta$ as the
coordinate of the dynamical system. We fix $\chi$ to be 3 for which two
SL states appear.
$f(\eta)$ is plotted in Fig. 5 as a function of $\eta$.  A and B are
the fixed points corresponding to the SL states. If $f(\eta) > 0$, the
flow of the dynamical variable is in the right direction, or else, in
the left direction, as indicated by arrows in the figure. It is clear from
the flow diagram that A is a stable fixed point, whereas B is unstable.
Therefore, of the two antisymmetric SL states one is stable and another
one is unstable. Similarly, we analyze the stability for all
the states. It is found that if $N$ states exist for a particular
case, $N$/2 states are stable provided that $N$ is even and,
otherwise ($N$+1)/2 states are stable.
\vspace{0.5cm}
     \begin{center}
      \epsfig{file=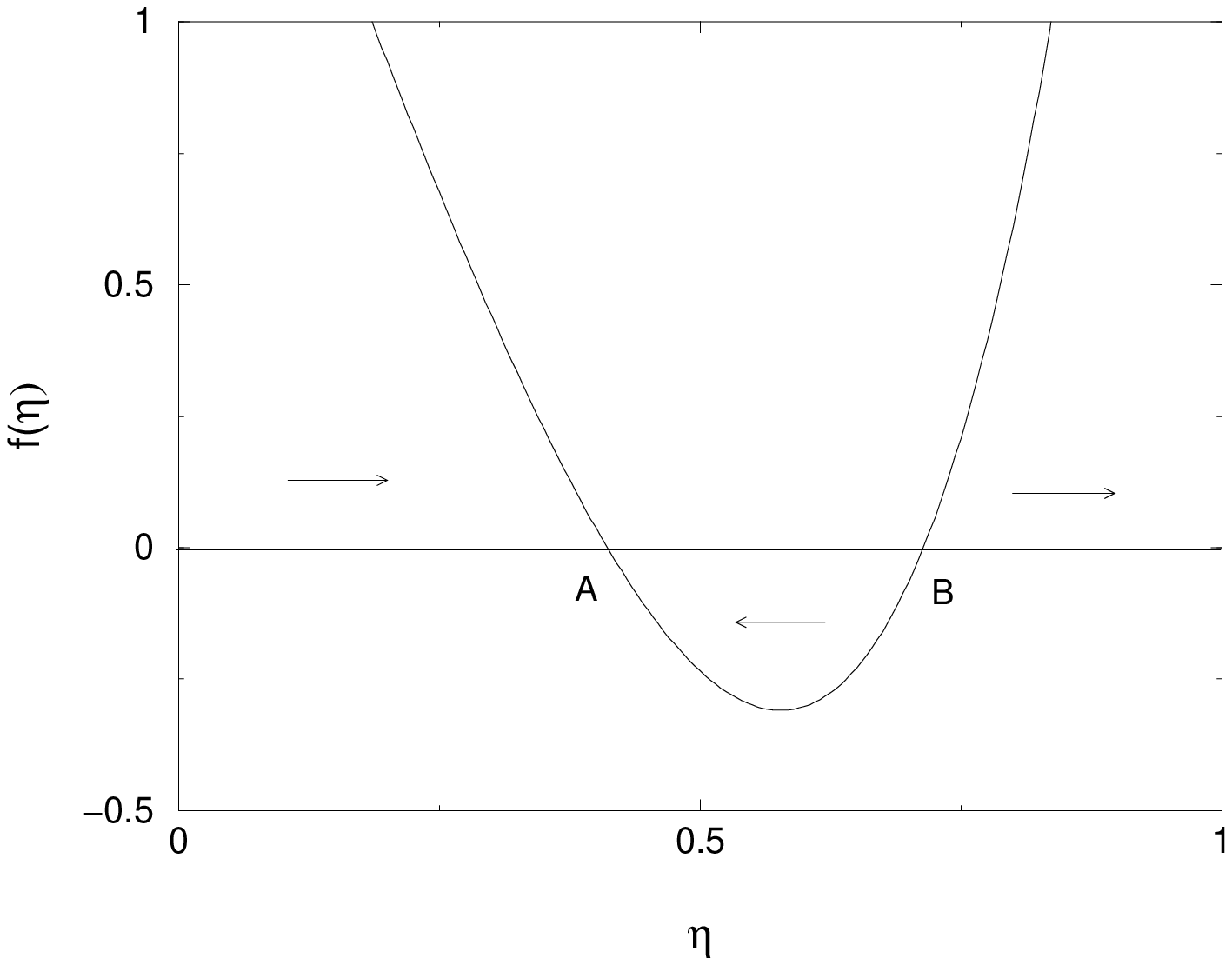,width=7.4cm,height=5.2cm}
      \vspace{0.4cm}
      \begin{figure}[p]
      \caption{The flow diagram where $f(\eta)$ is plotted as 
               a function of $\eta$. A and B are the fixed points. Point 
               A is a stable while B is unstable. 
      \label{fig5}}
      \vspace{0.5cm}
      \end{figure}
      \end{center}

\section{Conclusion}
Modified discrete nonlinear Schr\"odinger equation has been used to
study the formation of stationary localized states in a one-dimensional
system with a single nonlinear impurity and a dimer (symmetric as well 
as asymmetric)
nonlinear impurity. It is found that the critical values of nonlinear
strength decrease due to the nearest neighbor coupling of the lattice
oscillators.  Furthermore, a periodically modulated nonlinear chain is
studied and new SL states are found due to the modulation of nonlinearity.
It is important to note that there will be more number of SL states
if the periodicity of the  modulation is larger. Another interesting point
to note is that the critical value of nonlinearity to obtain SL state
decreases drastically due to the modulation of the nonlinearity in
a fully nonlinear chain and the asymmetric nature of the dimer impurity.
The phase diagrams of SL states are presented for all the cases. The mean
square displacement of a particle in one-dimensional chain with random
nonlinear site energies is calculated to investigate whether all the 
states are localized as it happens in the case of a linear
system. However, it is found that a fraction of the particle flows 
ballistically
indicating that all the states are not localized even in the presence
of random nonlinearities through out the system. It is further
found that SL states may appear even in a perfectly nonlinear system.
Thus, one may call these SL states as {\it self-localized} states.
The stability of the SL states is also discussed. To see the effect of
these nonlinearities on the formation of SL states in lattices with higher 
connectivity and higher dimension, the Cayley tree may be a good candidate
as a linear host lattice. 

\end{document}